# EPISODIC MAGNETIC BUBBLES AND JETS: ASTROPHYSICAL IMPLICATIONS FROM LABORATORY EXPERIMENTS

A. Ciardi[1*], S. V. Lebedev[2], A. Frank[3], F. Suzuki-Vidal[2], G. N. Hall[2], S. N. Bland[2], A. Harvey-Thompson[2], E. G. Blackman[3], M. Camenzind[4]

[1]*Observatoire de Paris, LERMA, 5 Place J Janssen, 92195 Meudon, France;*

[2]*Imperial College, Blackett Laboratory, Prince Consort Road, SW7 2BW, London, UK;* [3]*University of Rochester, Department of Physics and Astronomy, Rochester, NY, USA;*

[4] *University of Heidelberg, Centre for Astronomy Heidelberg (ZAH), Landessternwarte Koenigstuhl D-69117, Heidelberg, Germany.*

**ABSTRACT**

Collimated outflows (jets) are ubiquitous in the universe appearing around sources as diverse as protostars and extragalactic supermassive blackholes. Jets are thought to be magnetically collimated, and launched from a magnetized accretion disk surrounding a compact gravitating object. We have developed the first laboratory experiments to address time-dependent, episodic phenomena relevant to the poorly understood jet acceleration and collimation region. The experimental results show the periodic ejections of magnetic bubbles naturally evolving into a heterogeneous jet propagating inside a channel made of self-collimated magnetic cavities. The results provide a unique view of the possible transition from a relatively steady-state jet launching to the observed highly structured outflows.

**SUBJECT HEADINGS:** ISM: jets and outflows - - - ISM: Herbig-Haro objects - - -

*\*Present address: Ecole Normale Superieure, Laboratoire de Radioastronomie, Paris, France (andrea.ciardi@lra.ens.fr)*

**INTRODUCTION**

Although jets and outflows are associated with widely diverse astrophysical environments, they exhibit many common features independent of the central source (Livio 2002). In all contexts jets are believed to be driven by a combination of magnetic fields and rotation via, in most cases, an accretion disk (Lovelace 1976; Blandford & Payne 1982; Ferreira 1997; Ouyed et al. 1997). The standard magnetohydrodynamic jet models rely on rotation to twist a large-scale poloidal magnetic field $B_P$ and producing a toroidal field $B_\phi$ that drives and collimates the outflow. At some distance $d_L$ from the source, roughly the distance where the flow speed is Alfvénic, the continuous amplification of $B_\phi$ results in a region of "wound-up" field, $|B_\phi| \geq B_P$ where the outflow is collimated into a jet by the magnetic tension and further accelerated by gradients of the magnetic pressure. Our experiments are designed to model the acceleration and collimation of astrophysical jets taking place beyond $\sim d_L$, under the initial conditions $|B_\phi| \sim r^{-1} \gg B_P$, where $r$ is the radial distance from the axis. The experiments also apply to "magnetic tower" models (Lynden-Bell 1996, Nakamura et al. 2007), where differential rotation of closed magnetic field lines creates a highly wound magnetic field which drives a magnetic cavity as well as collimating a jet on its axis. Generally the distance $d_L$ and the amount of acceleration occurring before and after this distance is model dependent, and it is related to the unknown initial launching of the flow. For protostellar sources observations constrain jet launching to occur within a few AU from the source (Ray et al. 2007), and simulations indicate $d_L \sim 1-10$ AU (Zanni et al. 2007), which should be compared to jet propagation lengths of up to a few parsecs (Reipurth & Bally 2001). Indeed, how jets survive such unstable field configurations is one of the open astrophysical issues addressed by the experiments. Another aspect is the origin of the spatial and temporal variability that is observed on all scales in stellar jets (Hartigan et al. 2005), and which is often interpreted as perturbations to a relatively steady flow.

By performing appropriately scaled, high-energy density plasma experiments, "extreme" laboratory astrophysics has recently emerged as a novel approach to complement our understanding of complex astrophysical phenomena (see Remington et al. 2006 for a review). Jets have been the subject several studies, which may be distinguished (Blackman 2007) by whether they addressed problems related to the propagation (e,g Foster et al. 2005, Ciardi et al. 2008), the launching (Hsu & Bellan 2002), or both (Lebedev et al. 2005; Ciardi et al. 2007). Here we present the first laboratory experiments exploring episodic, magnetohydrodynamic (MHD) jets.

**ASTROPHYSICAL PHENOMENA IN THE LABORATORY**

Astrophysical jets and outflows are described to first approximation by ideal MHD and our experiments are designed to produce flows in that regime. Its applicability requires the dimensionless Reynolds ($Re$), magnetic Reynolds ($Re_M$) and Peclet ($Pe$) numbers to be much larger than unity; this implies that the transport of momentum, magnetic fields and thermal energy respectively, occurs predominantly through advection with the flow. It is important to stress that astrophysical jets have typical values $Re > 10^8$, $Re_M > 10^{15}$ and $Pe > 10^7$ that are *many orders of magnitude* greater than those obtained not only in the laboratory but also in numerical simulations, which have been so far the sole means of investigating time-dependent behaviour of multi-dimensional MHD jets. In ideal MHD simulations unphysical dissipation occurs at the grid level through numerical truncation errors (Ryu et al. 1995; Lesaffre & Balbus 2007). For global jet models, mostly performed assuming axisymmetry, the effective *numerical* Reynolds numbers are typically in the range $Re_M \sim 10 - 10^3$ (see for example Goodson & Winglee 1999), and we expect $Re \sim Pe \sim Re_M$. Severe limitations also exist on the range of plasma-$\beta$, the ratio of thermal to magnetic pressure, which may be reliably modelled numerically (Miller & Stone 2000). The (inherently) three-dimensional, scaled experiments discussed here extend the range of the dimensionless parameters obtained in the global modelling of jets. Typical values obtained are $Re \sim 5 \times 10^5 - 10^6$, $Re_M \sim 150 - 500$, $Pe \sim 20 - 50$ and $\beta \sim 10^{-3} - 10^3$; the plasma is highly collisional and the fluid MHD approximation is a valid model. Finally, similar to astrophysical jets, the laboratory flows produced are radiatively cooled.

**RESULTS**

The experiments were performed on the MAGPIE pulsed-power facility (current ~ 1 MA over 240 ns) and the associated simulations were carried out with our 3D resistive MHD code GORGON (Ciardi et al. 2007). A schematic diagram of the evolution of a typical experimental jet/outflow is shown in Figure 1. Two outflow components are generally present: a magnetic bubble (or cavity) accelerated by gradients of the magnetic pressure and surrounded by a shell of swept up ambient material, and a magnetically confined jet on the interior of the bubble. The confinement of the magnetic cavity itself relies on the presence of an external medium, thus confirming the theoretical picture of magnetic tower outflows (Lynden-Bell 1996).

The dynamics of the first magnetic bubble and jet are similar to those described by Lebedev et al. 2005; Ciardi et al. 2007. In the present work however we are able to produce and observe for the first time an episodic jet activity. The main difference with our previous experiments is an increased mass, as a function of radius, being available in the plasma source. The initial gap (cf. Figure 1) produced by the magnetic field is smaller and can be more easily refilled by the readily available plasma. Its closure allows the current to flow once again across the base of the magnetic cavity, thus re-establishing the initial configuration. When the magnetic pressure is large enough to break through this newly deposited mass, a new jet/bubble ejection cycle begins.

Figure 2 shows the observed correlation between x-ray bursts and the formation of a new bubble/jet system. The fast rising (~ 5 ns) part and peak of the emission are related to the maximum compression of the jet. We measured jet temperatures up to $\sim 3 \times 10^6$ K using spatially resolved, time-integrated spectroscopic measurements of H-like to He-like line ratios (Apruzese et al. 1997). Typical flow velocities observed are ~ 100 – 400 km s$^{-1}$, the simulated sonic and the alfvénic Mach numbers in the jet, defined as the ratios of the flow speed to the sound and Alfvén speed respectively, are $M_S \sim M_A \sim 3-10$. The characteristic energetics of the outflow can be estimated experimentally. The electromagnetic energy delivered to a typical bubble via Poynting flux is ~ 800 J and we estimate the kinetic energy of the flow to be ~ 100 - 400 J, the remainder is in the magnetic energy, internal energy of the plasma and partly lost to radiation. The time evolution of the jets and bubbles are presented in Figure 3. A succession of multiple cavities and embedded jets are seen to propagate over length scales spanning more than an order of magnitude. The resulting flow is heterogeneous and clumpy, and it is injected into a long lasting and well collimated channel made of nested cavities. It is worth remarking that the bow-shaped envelope is driven by the magnetic field and not hydrodynamically by the jet. The magnetically confined jet is in a configuration unstable to current-driven (CD) $m=0$ ("pinch") and non-axisymmetric $m=1$ ("kink") modes (Hardee 2004). The characteristic growth time is a few nano-seconds (Lebedev et al. 2005), and it is of the order of the Alfvén propagation time across the flow. For each ejection episode, lasting many e-folding times, the instabilities progress into their non-linear regime. In general the jets are seen to be going unstable via a CD kink mode which rather than disrupting the entire outflow saturates into a highly collimated super-alfvénic clumpy stream of "plasmoids". These clumps then continue to propagate ballistically detaching from the acceleration region (the source). Astrophysical jet simulations indicate that self-organizing processes may also limit the destructive nature of the Kelvin-Helmholtz instability (Ouyed et al. 2003). Thus jets may be formally unstable on relatively short time scales but these instabilities do not disrupt the overall long-term jet propagation. Our simulations show that the outflow has an "onion-like" density structure, and a collimation angle, with respect to the axial direction, of typically less ~ $30^o$ - $40^o$ for the low density plasma and less ~ $10^o$ - $20^o$ for the high density clumps. The instabilities also substantially modify the magnetic field. The kink mode leads to the generation of poloidal flux from the initially toroidal flux and to the tangling of the field inside the cavity and jet (Figure 4). Small-scale magnetic field structures develop which promote the dissipation of magnetic energy, ohmically and possibly via reconnection, and lead to further heating of the plasma. Moreover, magnetic energy dissipation may also cause additional acceleration of the bulk plasma (Drenkhahn & Spruit 2002). Each jet/outflow episode propagates, interacts and substantially alters the surrounding environment by injecting mass, momentum, energy and magnetic flux into it. An important aspect of the episodic ejection process is, broadly speaking, its self-collimation. Since the initial ambient medium is swept away after a few ejections, newly formed magnetic cavities are confined solely by the environment left by earlier episodes, thus making the collimation process insensitive to the initial ambient conditions. In the magnetic cavities $\mathrm{Re}_M > 100$, and each bubble expands with its own "frozen-in" magnetic flux; in the experiments this is confirmed by the magnetic probe measurements of the trapped magnetic field at the outer edge of the bubbles, $B \sim 1-5 \text{ kG}$. The collimation is then determined not only by the pressure of the left-over plasma but also by the pressure of the tangled magnetic field trapped in the bubbles, where the plasma-$\beta$ is in the range $0.1 < \beta \sim 1$. A high level of symmetry is maintained after many ejections (~ 5), the number being limited only by the duration of the current pulse delivered by the generator. Overall, the experiments demonstrate that magnetic acceleration and collimation, occurring within a framework of strongly episodic outflow activity, can be effective in producing well collimated and heterogeneous jets.

**ASTROPHYSICAL IMPLICATIONS**

Two time-scales are of interest in the experiments and they may be related to astrophysical jet sources. The first is the growth time $\tau_I$ of the CD instabilities, and for conditions applicable to the formation region of protostellar jets (Hartigan et al. 2007) we can estimate the growth time of the CD kink mode as the Alfvén crossing time $\tau_I \sim 1 \text{ year}$; corresponding to a few nanoseconds in the experiments. The second time-scale is the relatively longer bubble ejection period $\tau_B$, which is linked to the temporal variability of the Poynting flux feeding the bubbles. For astrophysical sources $\tau_B$ should be associated with a substantial variation in the outflow launching activity. Although the mechanism responsible is unknown, observations of knots kinematics, assumed to trace ejection variability, suggest characteristic times of $5-20 \text{ years}$, and we take the bubble ejection period $\tau_B$ to be of the same order; the experiments are in a similar regime $\tau_B/\tau_I \sim 10$. We stress that both time-scales $\tau_B$ and $\tau_I$ are relatively longer than the characteristic Keplerian period of rotation at the inner disk radius, and in this respect jet launching should have ample time to reach steady-state. Taking the characteristic astrophysical flow velocities to be $v \sim 200 \text{ km s}^{-1}$, which is of the same order as the experiments and does not need scaling, we can draw a parallel between protostellar outflows and the dynamics observed in the experiments. The presence of multiple bubble-like features should be observed on scales $L \sim \tau_B v$ ranging from a few tens to a few hundred AU from the source. Because of the relatively short growth time of the instabilities $\tau_I < \tau_B$, jets should develop non-axisymmetric features already within a few tens AU from the source, and become more heterogeneous and clumpy as they move further away to hundreds of AU. Over the same length scales we would expect magnetic energy dissipation, heating of the plasma and a transition to a kinetically dominated jet which propagates ballistically. A compelling astrophysical system that is related to the picture emerging from our experiments is the outflow from the TTauri star DG Tau. Ejection variability, limb-brightened bubble-like structures and the presence of wiggles in the optical jet are indeed evident on scales ranging from of a few tens to a few hundred AU the source (Bacciotti et al. 2000; Dougados et al. 2000). It was also recently reported for a number of TTauri jets including DG Tau, that already within 100 AU from the source the jet physical conditions show considerable asymmetries with respect to the axis (Coffey et al. 2008), the experiments indicate that asymmetries in the flow can be produced by instabilities that do not destroy the collimation. X-ray emission from the DG Tau jet was also recently detected on the same length scales and it was suggested that magnetic energy dissipation may be behind the heating mechanism (Güdel et al. 2008). As in the experiments, instabilities and the tangling of the field may provide a compelling route to efficient heating of such plasmas. Finally we note that XZ Tau shows episodic outflow activity (Coffey et al. 2004) in the form of rapidly evolving "bubbles" with a clumpy jet-like morphology along the axis (Krist et al. 1999).

We presented the first experimental *simulation* of episodic, astrophysical jet launching. The inherent three-dimensionality of the experiments, coupled with large dimensionless numbers, radiative cooling, and a wide range of plasma-$\beta$, make them an important new tool to understand the evolution of astrophysical jets. Two key principles emerge: first, toroidal fields can collimate and accelerate flows to super-magnetosonic speeds. Second, even if the average flow geometry and collimation is "steady" over long time-scales, the jet activity can be episodic and the instabilities need not disrupt the long term collimation. The results offer a first glimpse on how magnetohydrodynamic outflows may naturally evolve from a relatively steady-state launching to a heterogeneous jet, suggesting a new scenario which shifts emphasis away from stationary, steady-state production of continuous jet beams, towards an episodic, ejection of plasmoids surrounded by evolving field configurations.

**Acknowledgements**

This work was supported in part by the European Community JETSET network, RTN-CT-2004 005592. We acknowledge the London e-Science Centre for the provision of computational facilities.

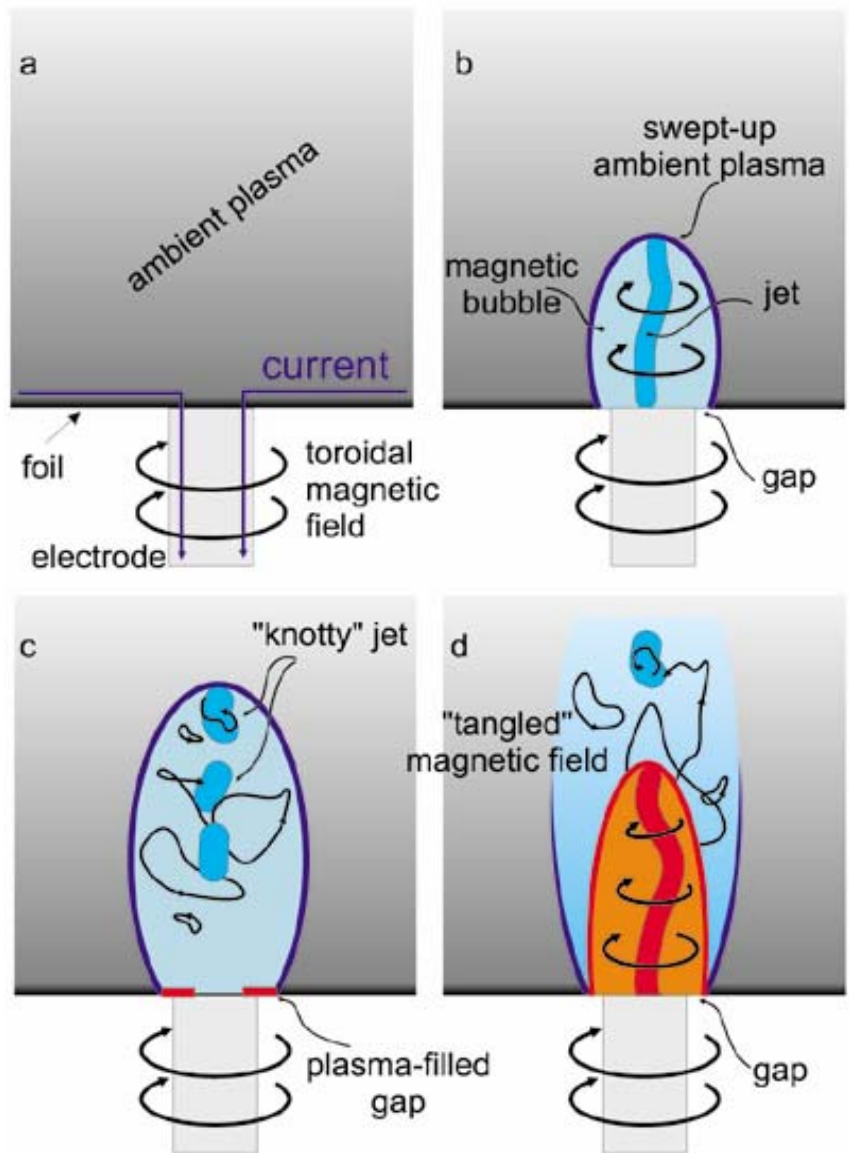

FIGURE 1

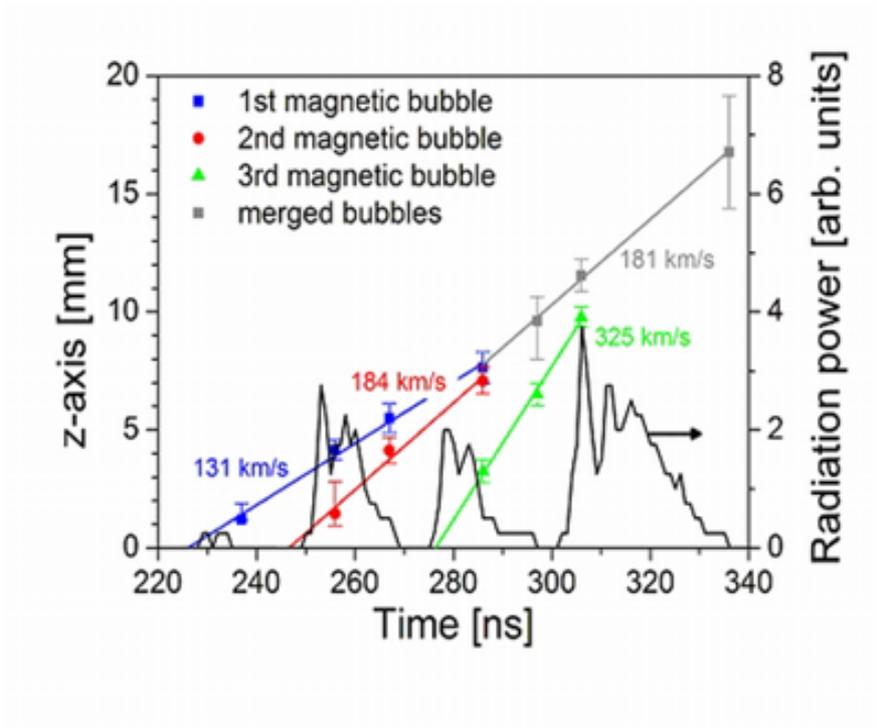

FIGURE 2

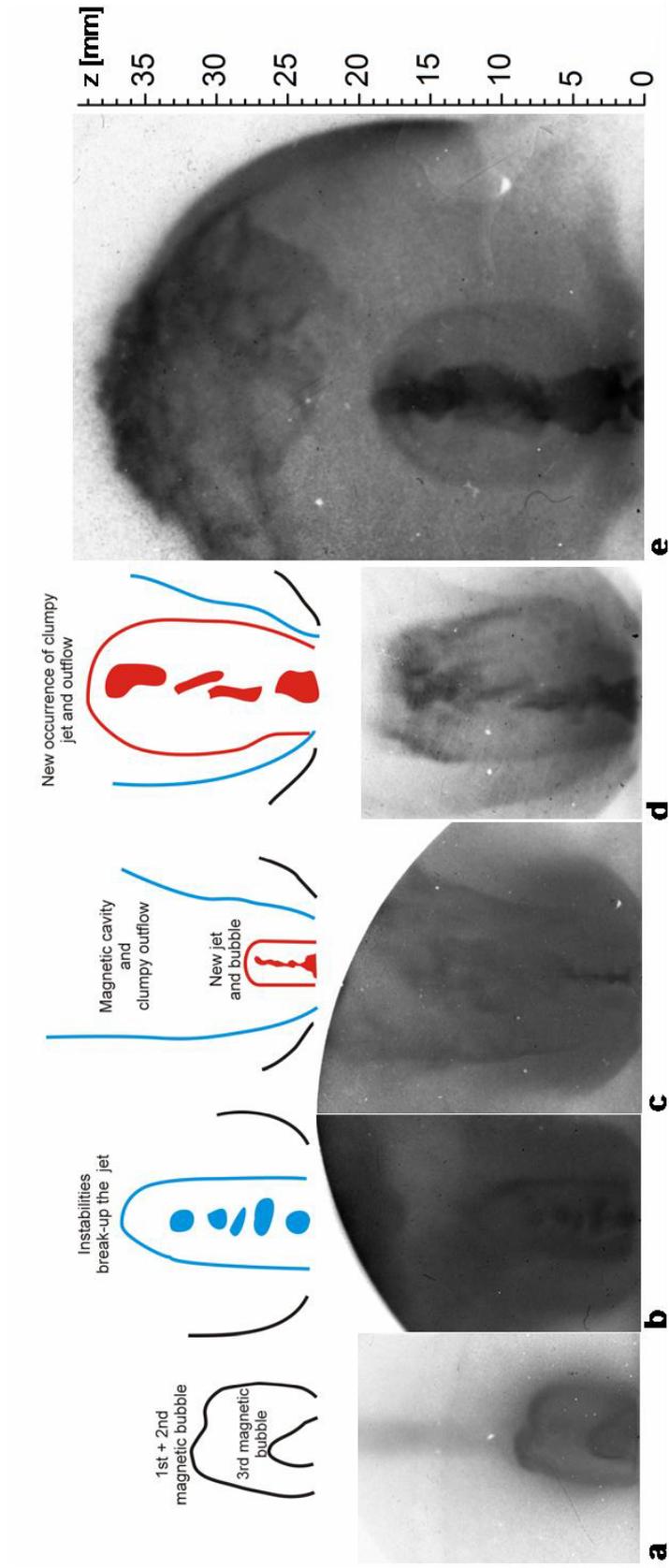

FIGURE 3

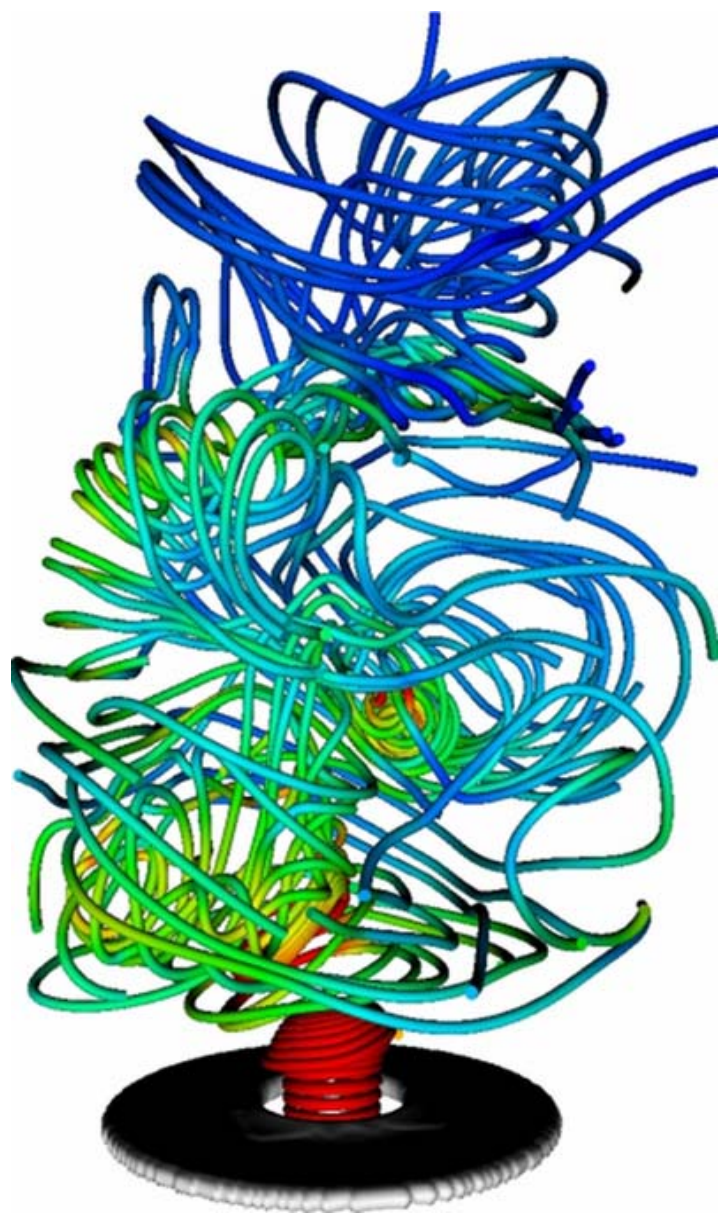

FIGURE 4

Figure 1 Schematic evolution of a jet/bubble system. The experimental load consists of 6 μm thick Aluminium foil connecting two co-axial electrodes. (a) During the first ~ 200 ns (times are given from the start of the current pulse) the ohmically heated foil produces an ambient plasma extending a few millimetres above the foil. (b) At ~ 230 ns the accumulated action of the magnetic pressure breaks the foil near the central electrode, effectively creating a "gap" (~ 0.5 – 1 mm) between the cathode and the left-over foil. The magnetic field then accelerates and sweeps the ambient plasma into a shock-layer which confines and delineates the bubble. At the same time a current-carrying jet forms on the axis of the bubble via the "pinching" effect of the toroidal magnetic field. (c) The magnetized jet is unstable and the flow is re-arranged into a collimated "knotty" jet. The "gap" at the base of the bubble is refilled by plasma expanding from the foil, and the initial configuration is restored. A new jet and bubble (shown in red) are then formed (d) which propagate and interact with the plasma and "tangled" magnetic field left by the previous ejection event (shown in blue). Again the "gap" is refilled with plasma and a new jet/bubble ejection cycle begins (not shown).

Figure 2 The radiation power obtained from filtered PCD shows the presence of four distinct "bumps" each corresponding to the formation of a magnetic bubble and jet. From time-resolved XUV emission images (obtained during one single experiment), the axial position of the top of each magnetic bubble is measured as a function of time. A linear fit shows the approximate axial expansion speed for each of the three separate bubbles and at later times for the merged bubbles. Shot [s0301].

Figure 3 Time-series of filtered XUV emission images taken at (a) 286 ns, (b) 346 ns, (c) 376 ns, (d) 406 ns, (e) 487 ns. The schematic cartoon serves to guide the eye; the features are described in the main text. The column-like emission present above the bubbles comes from plasma that has converged on axis prior to their formation (Ciardi et al. 2007) and it is of no interest here. Three different experimental shots are shown: (a) [s0301]; (b), (c) and (d) [s0530]; (e) [s0612].

Figure 4 3D MHD simulations of the experiments. Simulated magnetic field lines are shown at 305 ns; the colours represent the field strength decreasing from *red* to *blue*. The disk at the bottom is an isodensity surface showing the left-over foil. At the time shown, a third newly formed bubble can be seen emerging through the gap in the foil.